\documentclass[11pt,english]{article}
\usepackage[T1]{fontenc}
\usepackage[latin9]{inputenc}
\usepackage{geometry}
\usepackage{float}
\usepackage{url}
\usepackage{amsthm}
\usepackage{amsmath}
\usepackage{graphicx}
\usepackage{amssymb}

\floatstyle{ruled}
\newfloat{algorithm}{tbp}{loa}
\floatname{algorithm}{Algorithm}

\theoremstyle{plain}
\theoremstyle{plain}
\newtheorem{thm}{Theorem}
  \theoremstyle{plain}
  \newtheorem{lem}[thm]{Lemma}

\usepackage{babel}

\begin{document}

\title{A Deterministic Polynomial--Time Algorithm for Constructing a Multicast
Coding Scheme for Linear Deterministic Relay Networks}

\date{S. M. Sadegh Tabatabaei Yazdi and Serap A. Savari%
\thanks{The authors are with the Department of Electrical and Computer Engineering,
Texas A\&M University, College Station, TX 77843 USA. Their e-mail
addresses are \protect\url{sadegh@neo.tamu.edu} and \protect\url{savari@ece.tamu.edu}.%
}}
\maketitle
\begin{abstract}
We propose a new way to construct a multicast coding scheme for linear
deterministic relay networks. Our construction can be regarded as
a generalization of the well-known multicast network coding scheme
of Jaggi et al. to linear deterministic relay networks and is based
on the notion of flow for a unicast session that was introduced by
the authors in earlier work. We present randomized and deterministic
polynomial--time versions of our algorithm and show that for a network
with $g$ destinations, our deterministic algorithm can achieve the
capacity in $\left\lceil \log(g+1)\right\rceil $ uses of the network. 
\end{abstract}

\section{Introduction}

Computing the capacity and constructing optimal coding schemes for
wireless Gaussian networks are central open questions and of great
importance in network information theory. In a wireless network the
transmitted signal from a node is broadcasted to all its neighbors
and the signal received at a node is the superposition of the signals
transmitted by its neighbors and Gaussian noise. Broadcasting, interference,
and noise are the three main characteristics of a wireless network
that differentiate it from a wired network and make its analysis much
more challenging. Recently Avestimehr, Diggavi, and Tse \cite{avestimehr_1}
proposed an approximation model known as the \emph{linear deterministic
relay network} (LDRN) for wireless Gaussian networks that simplifies
the three features of wireless Gaussian networks by instead considering
deterministic and linear operations in vector spaces over finite fields.
Avestimehr, Diggavi, and Tse have further shown that for some Gaussian
wireless networks, the capacity of the wireless network is within
an additive constant gap from the capacity of the corresponding approximation
network and the optimal coding scheme for the approximation network
can be translated to near optimal coding schemes for the Gaussian
wireless network \cite{avestimehr_2}. 

\begin{figure}
\centering\includegraphics[bb=0bp 0bp 500bp 250bp,clip,scale=0.5]{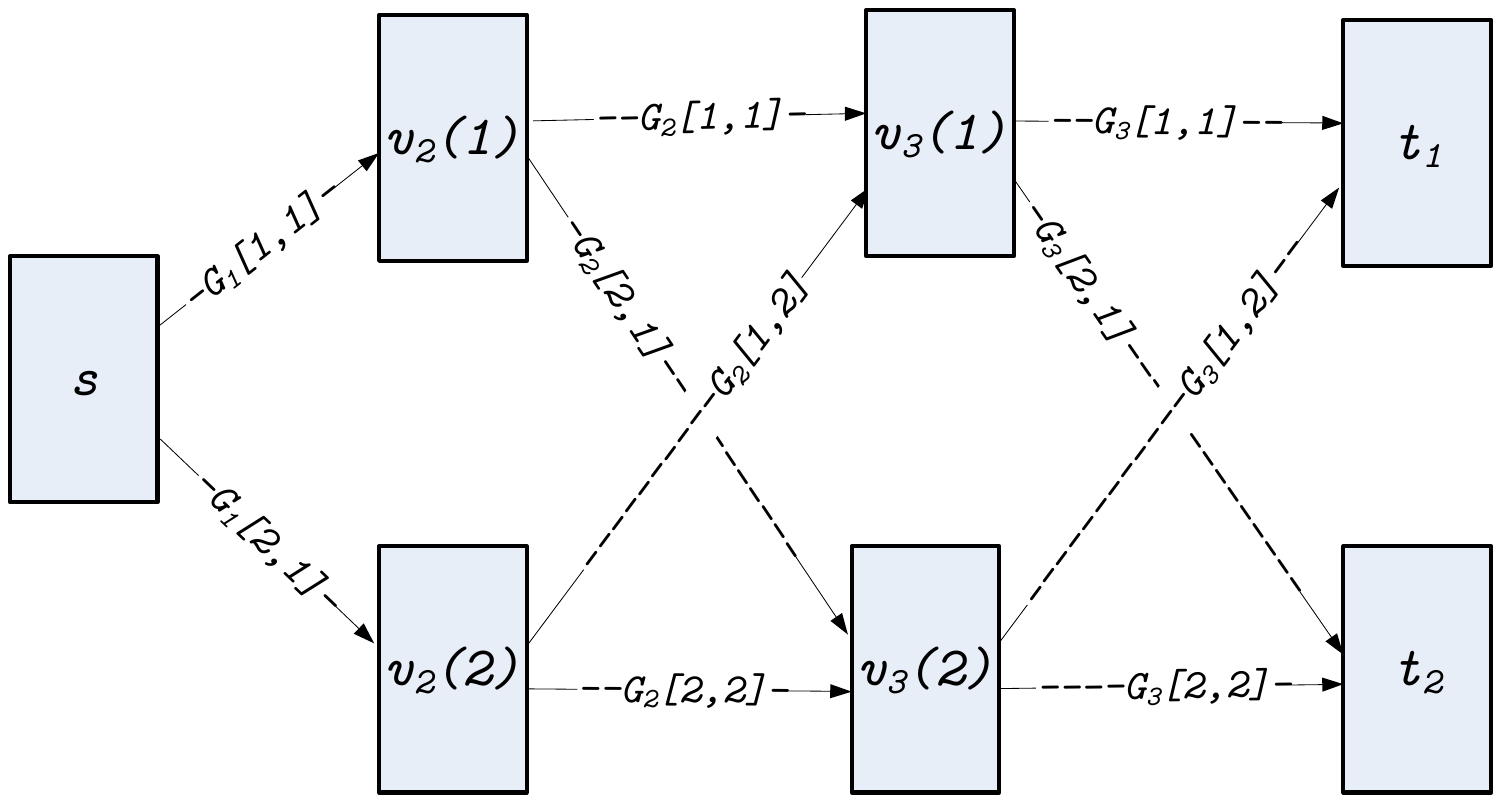}

\caption{\label{fig:An-LDRN-with}An LDRN with four layers. Here $t_{1}=v_{4}(1)$
and $t_{2}=v_{4}(2)$.}

\end{figure}

An LDRN is a wireless networking model which can be visualized as
a layered directed network $\mathcal{N}=(V,E)$ with set of {}``nodes''
$V=\bigcup_{i=1}^{M}V_{i}$, where $V_{i}$ denotes the set of nodes
in layer $i$, and set of {}``edges'' $E$. Let $V_{i}=\left\{ v_{i}(1),\cdots,v_{i}(m_{i})\right\} $,
where $m_{i}$ denotes the number of nodes in layer $i$. The first
layer consists of a single node $s=v_{1}(1)$ called the source node.
There are $g$ destination nodes denoted by $t_{l}\triangleq v_{K_{l}}(d_{l}),l\in\left\{ 1,\cdots,g\right\} $,
distributed in layers $K_{1},K_{2},\cdots,K_{g}$. There is an {}``edge''
from every node in $V_{i}$ to every node in $V_{i+1}$ which corresponds
to the \emph{transfer matrix} between the two nodes. Figure \ref{fig:An-LDRN-with}
is an example of an LDRN with four layers and two destination nodes.%

During one use of the communication channel between layers $i$ and
$i+1$, $v_{i}(j)$ transmits a predetermined length vector $\mathbf{x}_{i}[j]$
to the nodes in layer $i+1$ and $v_{i+1}(k)$ receives a predetermined
length vector $\mathbf{y}_{i+1}[k]$ given by \[
\mathbf{y}_{i+1}[k]=\sum_{j=1}^{m_{i}}G_{i}[k,j]\mathbf{x}_{i}[j],\]
 where $G_{i}[k,j]$ is a predetermined transfer matrix of the edge
$(v_{i}(j),v_{i+1}(k))\in E$. Note that we can set $G_{i}[k,j]$
to be the all-zero matrix if there is no connection from $v_{i}(j)$
to $v_{i+1}(k)$. All vectors and matrices are over a fixed finite
field ${\mathbb{F}}$. One can define\[
\mathbf{x}_{i}=\left[\begin{array}{c}
\mathbf{x}_{i}[1]\\
\vdots\\
\mathbf{x}_{i}[m_{i}]\end{array}\right],\mathbf{y}_{i+1}=\left[\begin{array}{c}
\mathbf{y}_{i+1}[1]\\
\vdots\\
\mathbf{y}_{i+1}[m_{i+1}]\end{array}\right]\]
and the block matrix $G_{i}=\left[G_{i}[k,j]\right],1\leq k\leq m_{i+1},1\leq j\leq m_{i}$.
Then the received vectors at layer $i+1$ are related to the transmitted
vectors at layer $i$ by the following relationship\[
\mathbf{y}_{i+1}=G_{i}\mathbf{x}_{i}.\]
 The capacity of an LDRN for a single multicast session from source
$s$ to the destinations $t_{1},\cdots,t_{g}$ is derived in \cite{avestimehr_1}.
Define a cut between the source node $s$ and a destination node $t_{j}$
as a partition of nodes $V$ into two sets $A$ and $B$, with $s\in A$
and $t_{j}\in B.$ The capacity of the cut is defined as the rank
of the transfer matrix from the transmitted vectors of the nodes in
$A$ to the received vectors of the nodes in $B.$ \cite{avestimehr_1}
shows that the minimum capacity of the cuts between $s$ and $t_{j}$
is the capacity of a unicast session between $s$ and $t_{j}.$ Furthermore
the multicast capacity of the network between source $s$ and destinations
$t_{1},\cdots,t_{g}$ is the minimum of the min--cut capacities between
the source and each destination. The capacity--achieving scheme in
\cite{avestimehr_1} is a random linear coding scheme that is asymptotically
optimal when the network is used for multiple rounds. 

A few groups of researchers (see, e.g., \cite{amaudruz,goemans,tabatabaei_1,tabatabaei_2})
have proposed deterministic coding schemes for the transmission of
a single unicast session over an LDRN which can be constructed in
polynomial time. Furthermore, they achieve capacity using only one
round of the network. These schemes are similar to routing schemes
in wired networks and have low encoding and decoding complexities
at the relay nodes. 

In this paper we build upon our work in \cite{tabatabaei_1,tabatabaei_2}
to design a simple and low complexity transmission scheme for a multicast
session over an LDRN. Our scheme will be constructed by progressively
combining the coding schemes for unicast sessions from the source
to each destination. In many ways our scheme is similar to and is
a generalization of the scheme in \cite{jaggi} for a multicast session
in wired networks. We will offer both randomized and deterministic
versions of our algorithm and show that $\left\lceil \log(g+1)\right\rceil $
uses of the network suffice to achieve capacity, which resembles the
result for wired networks \cite{jaggi}.

For the case of a single multicast session, there have been multiple
recent attempts to devise deterministic and efficient algorithms for
constructing capacity--achieving coding schemes. In \cite{ebrahimi},
Ebrahimi and Fragouli developed an algebraic framework for vector
network coding and used this framework to devise a multicast transmission
scheme over an LDRN. Our scheme has a lower complexity of construction
and needs fewer uses of the network to achieve capacity. Erez et al.
\cite{erez} offer a different construction by progressing through
the network according to a topological order and maintaining the linear
independence of certain subsets of coding vectors along the processing.
However, the proposed algorithm does not appear to have a polynomial
running time. Kim and Médard \cite{kim} generalized the algebraic
framework of Koetter and Médard \cite{koetter_medard} for classical
network coding to LDRNs and devised an algebraic algorithm for constructing
multicast codes. Again, the proposed algorithm does not appear to
have a polynomial running time. More recently, \cite{khojastepour}
proposed an algorithm using rotational codes to asymptotically achieve
the multicast capacity of LDRN networks for a multicast session. Rotational
codes have some built--in advantages as they are easy to implement
at the relay nodes. However, the existence of deterministic polynomial--time
algorithms for the construction of efficient rotational codes for
multicast transmission over an LDRN remains unknown.

We will next review our earlier results on a single unicast session
\cite{tabatabaei_1,tabatabaei_2} in Section \ref{sec:A-single-unicast}
and then discuss our coding construction for a multicast session in
Section \ref{sec:A-coding-scheme}.

\section{\label{sec:A-single-unicast}A single unicast session}

In this section we briefly explain the coding scheme for a single
unicast session from \cite{tabatabaei_1,tabatabaei_2}. This will
be the building block of our multicast coding scheme.

Recall that for each $i\in\left\{ 1,\cdots,M-1\right\} $ the transmitted
vector of layer $i$ and the received vector of layer $i+1$ are related
through matrix $G_{i}$ by $\mathbf{y}_{i+1}=G_{i}\mathbf{x}_{i}$. 

For each layer $i\in\left\{ 1,\cdots,M\right\} $ label the indices
of the vector $\mathbf{y}_{i}$ with the elements of a set $P_{i}$
and label the indices of the vector $\mathbf{x}_{i}$ with the elements
of a set $Q_{i}.$ We choose all sets $P_{i}$ and $Q_{i}$ to be
disjoint for different values of $i$. For any $A\subseteq P_{i},$
let $\mathbf{y}_{i}(A)$ denote the subvector of $\mathbf{y}_{i}$
corresponding to indices with labels from set $A.$ Similarly, for
any $B\subseteq Q_{i},$ let $\mathbf{x}_{i}(B)$ denote the subvector
of $\mathbf{x}_{i}$ associated with indices with labels from set
$B.$ Next partition each set $P_{i}$ into subset $P_{i}=\cup_{j=1}^{m_{i}}P_{i}[j]$
and $Q_{i}$ into subsets $Q_{i}=\cup_{j=1}^{m_{i}}Q_{i}[j]$ such
that $P_{i}[j]$ is the subset of indices of $\mathbf{y}_{i}$ that
belong to the subvector $\mathbf{y}_{i}[j]$ and $Q_{i}[j]$ is the
subset of indices of $\mathbf{x}_{i}$ that belong to the subvector
$\mathbf{x}_{i}[j]$. Therefore we have $\mathbf{y}_{i}[j]=\mathbf{y}_{i}(P_{i}[j])$
and $\mathbf{x}_{i}[j]=\mathbf{x}_{i}(Q_{i}[j])$ for any $j\in\left\{ 1,\cdots,m_{i}\right\} .$
For any $i\in\left\{ 1,\cdots,M-1\right\} $ we will use the sets
$P_{i+1}$ and $Q_{i}$ to label the rows and the columns of the matrix
$G_{i}$ such that for each $p\in P_{i+1}$ the row of $G_{i}$ corresponding
to the element $\mathbf{y}_{i+1}(p)$ is labeled with $p$ and for
each $q\in Q_{i}$ the column of $G_{i}$ corresponding to the element
$\mathbf{x}_{i}(q)$ is labeled with $q$. For $p\in P_{i+1}$ and
$q\in Q_{i}$ let $G_{i}(p,q)$ denote the element in row $p$ and
column $q$ of matrix $G_{i}.$ For $A\subseteq P_{i+1}$ and $B\subseteq Q_{i}$
let $G_{i}(A,B)$ denote the submatrix of $G_{i}$ consisting of the
rows in $A$ and the columns in $B.$ Our labeling implies that $G_{i}(P_{i+1}[k],Q_{i}[j])=G_{i}[k,j]$
for any $j\in\left\{ 1,\cdots,m_{i}\right\} $ and $k\in\left\{ 1,\cdots,m_{i+1}\right\} .$

If node $s$ holds a column vector message $\mathbf{w}\in\mathbb{F}^{R\times1}$
and we are looking at a linear coding scheme, then at each layer $i\in\left\{ 1,\cdots,M\right\} ,$
each element of vectors $\mathbf{x}_{i}$ and $\mathbf{y}_{i}$ will
be a linear transformation of the vector $\mathbf{w}.$ We represent
the {}``global coding vector'' (see \cite{jaggi}) for the element
$\mathbf{x}_{i}(q),q\in Q_{i},$ with row vector $\boldsymbol{x}_{i}(q)\in\mathbb{F}^{1\times R}$
such that $\mathbf{x}_{i}(q)=\boldsymbol{x}_{i}(q)\mathbf{w}$ and
the global coding vector for the element $\mathbf{y}_{i}(p),p\in P_{i},$
with row vector $\boldsymbol{y}_{i}(p)\in\mathbb{F}^{1\times R}$
such that $\mathbf{y}_{i}(p)=\boldsymbol{y}_{i}(p)\mathbf{w}$. For
subsets $B\subseteq Q_{i}$ and $A\subseteq P_{i}$ we use the notation
$\boldsymbol{x}_{i}(B)$ and $\boldsymbol{y}_{i}(A)$ to respectively
denote the matrices that are formed by the vectors $\boldsymbol{x}_{i}(q)$
and $\boldsymbol{y}_{i}(p)$ for $q\in B$ and $p\in A$. Therefore
we have $\mathbf{x}_{i}(B)=\boldsymbol{x}_{i}(B)\mathbf{w}$ and $\mathbf{y}_{i}(A)=\boldsymbol{y}_{i}(A)\mathbf{w}$.

Suppose that the network supports a rate--$R$ unicast connection
between source node $s$ and a destination node $t=v_{K}(d)$ for
$K\leq M$ and $d\in\left\{ 1,\cdots,m_{K}\right\} .$ The main result
of \cite{tabatabaei_1,tabatabaei_2} can be summarized in the following
theorem:
\begin{thm}
\label{thm:unicastflow}For each $1\leq i\leq K$ and for each $1\leq j,k\leq m_{i}$
there exist subsets $\hat{Q}_{i}[j]\subseteq Q_{i}[j]$ and $\hat{P}_{i}[k]\subseteq P_{i}[k]$
such that the following hold
\begin{enumerate}
\item $|\hat{P}_{i}[j]|=|\hat{Q}_{i}[j]|$ for $i\in\left\{ 1,\cdots,K\right\} ,j\in\left\{ 1,\cdots,m_{i}\right\} ,$
\item $\sum_{j=1}^{m_{i}}|\hat{P}_{i}[j]|=\sum_{j=1}^{m_{i}}|\hat{Q}_{i}[j]|=R,$
for $i\in\left\{ 1,\cdots,K-1\right\} ,$
\item $|\hat{P}_{K}[d]|=R$ and $|\hat{P}_{K}[k]|=0$ for $k\neq d,$
\item $G_{i}(\bigcup_{k=1}^{m_{i+1}}\hat{P}_{i+1}[k],\bigcup_{j=1}^{m_{i}}\hat{Q}_{i}[j])$
is a nonsingular matrix for $i\in\left\{ 1,\cdots,K-1\right\} .$ 
\end{enumerate}
Furthermore such subsets can be found by an algorithm that runs in
a time that is polynomial in the size of the network $\mathcal{N}.$
\end{thm}
We call the subsets $\hat{Q}_{i}[j]\subseteq Q_{i}[j]$ and $\hat{P}_{i}[k]\subseteq P_{i}[k]$
for $i\in\left\{ 1,\cdots,M\right\} $ and $j,k\in\left\{ 1,\cdots,m_{i}\right\} $
a \emph{flow} of rate $R$ in the LDRN from the source node $s$ to
the destination node $t.$ 

The four properties of a flow in Theorem \ref{thm:unicastflow} depend
on $G_{1},...,G_{M-1}$ and do not depend on the specific choice of
the set $\hat{P}_{1}[1]$ among all subsets of $P_{1}[1]$ with size
$R.$ Therefore, if there exists a rate--$R$ flow, we can set $\hat{P}_{1}[1]$
to be any subset of $P_{1}[1]$ of size $R.$ %
\begin{figure}
\centering\includegraphics[bb=0bp 0bp 500bp 250bp,clip,scale=0.5]{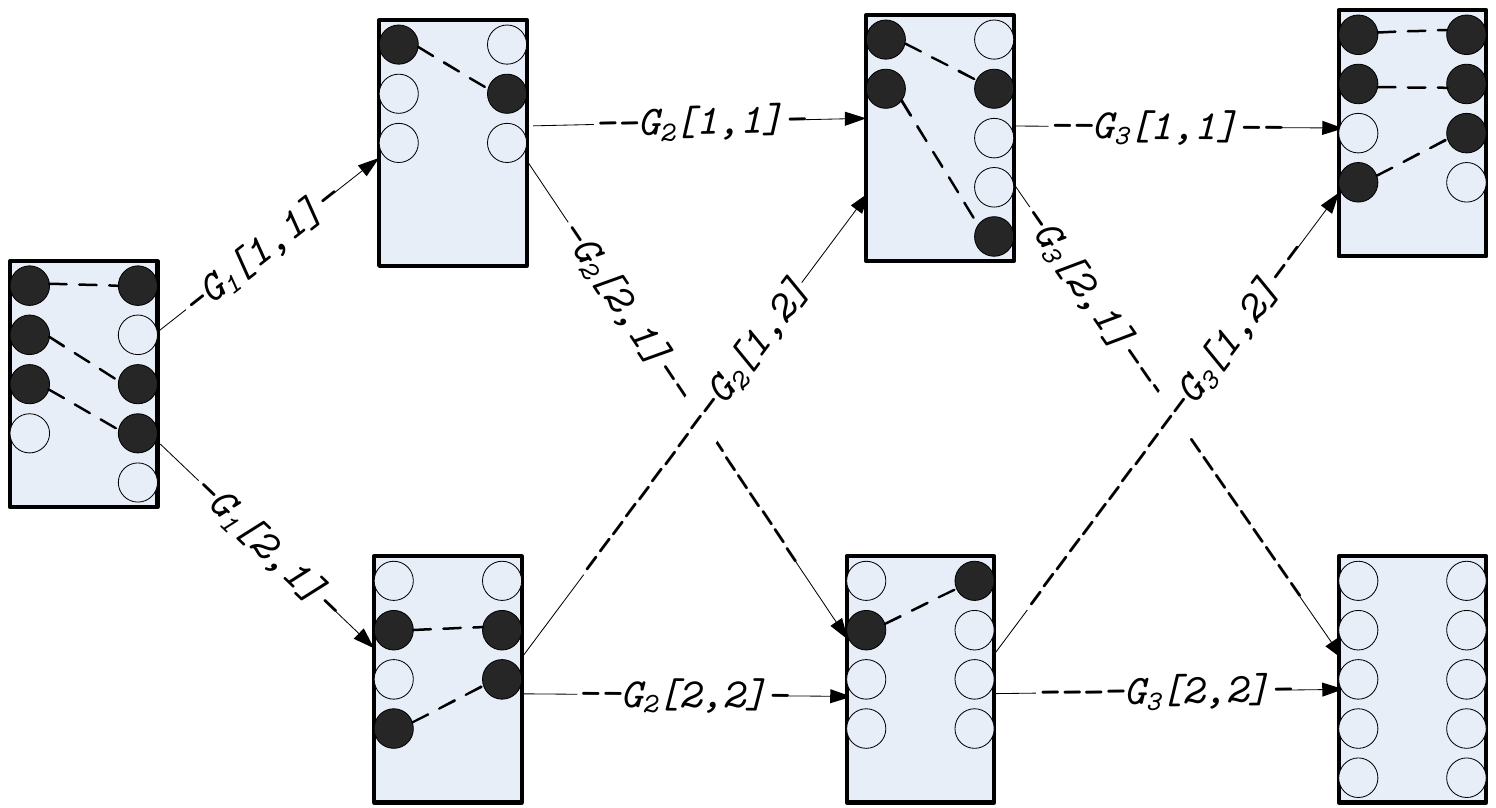}

\caption{\label{fig:example_of_flow}An example of a rate--3 flow from the
source node $s$ to the destination node $t_{1}$. Here the matched
elements of flow are connected together through dashed lines.}

\end{figure}

 Notice that the existence of a flow of rate $R$ implies the following
simple and low complexity coding scheme of rate $R$ from the source
$s$ to the destination $t$: To send message $\mathbf{w}\in\mathbb{F}^{R\times1},$
source node $s=v_{1}(1)$ sets $\mathbf{y}_{1}(\hat{P}_{1}[1])=\mathbf{w}$
and $\mathbf{y}_{1}(P_{1}[1]\setminus\hat{P}_{1}[1])=\mathbf{0}.$
Next, any node $v_{i}(j),i\in\left\{ 1,\cdots,M\right\} ,j\in\left\{ 1,\cdots,m_{i}\right\} ,$
in the network forms the vector $\mathbf{x}_{i}[j]$ by setting \[
\mathbf{x}_{i}(\hat{Q}_{i}[j])=\mathbf{y}_{i}(\hat{P}_{i}[j]).\]
We say that element $p\in\hat{P}_{i}[j]$ is {}``matched'' with
element $q\in\hat{Q}_{i}[j]$ when $\mathbf{x}_{i}(q)$ is set to
$\mathbf{y}_{i}(p)$ through the preceding equation (see Figure \ref{fig:example_of_flow}
for an example of a flow). We further let $\mathbf{x}_{i}(Q_{i}[j]\setminus\hat{Q}_{i}[j])=\mathbf{0}.$
From the properties of flow it follows that at the destination $t=v_{K}(d)$
\[
\mathbf{x}_{K}(\hat{Q}_{K}[d])=G_{K-1}(\hat{P}_{K}[d],\bigcup_{j=1}^{m_{K-1}}\hat{Q}_{K-1}[j])\cdots G_{2}(\bigcup_{k=1}^{m_{3}}\hat{P}_{3}[k],\bigcup_{j=1}^{m_{2}}\hat{Q}_{2}[j])G_{1}(\bigcup_{k=1}^{m_{2}}\hat{P}_{2}[k],\hat{Q}_{1}[1])\mathbf{w}.\]
 Since each matrix $G_{i}(\bigcup_{k=1}^{m_{i+1}}\hat{P}_{i+1}[k],\bigcup_{j=1}^{m_{i}}\hat{Q}_{i}[j])$
is nonsingular, node $t$ can recover vector $\mathbf{w}$ from the
received vector $\mathbf{x}_{K}(\hat{Q}_{K}[d])$ through a linear
transformation.

\section{\label{sec:A-coding-scheme}A coding scheme for a multicast session }

Assume that there are $g$ destination nodes $t_{1},\cdots,t_{g}$
in the network and the min--cut capacity from the source node $s$
to each destination is at least $R.$ We are interested in a multicast
coding scheme in which all destinations can simultaneously receive
the message $\mathbf{w}\in\mathbb{F}^{R\times1}$ of the source. Our
scheme will be designed by combining the flows of rate $R$ from the
source to each destination. 

Suppose that $t_{l}=v_{K_{l}}(d_{l})$ for $l\in\left\{ 1,\cdots,g\right\} .$
From Theorem \ref{thm:unicastflow} for each $t_{l},l\in\left\{ 1,\cdots,g\right\} $,
there exists a flow with subsets $P_{i}^{l}[k]\subseteq P_{i}[k]$
and $Q_{i}^{l}[j]\subseteq Q_{i}[j]$ for $1\leq i\leq K_{l}$ and
$1\leq j,k\leq m_{i}$ such that:
\begin{enumerate}
\item $|P_{i}^{l}[j]|=|Q_{i}^{l}[j]|$ for $i\in\left\{ 1,\cdots,K_{l}\right\} ,j\in\left\{ 1,\cdots,m_{i}\right\} ,$
\item $\sum_{j=1}^{m_{i}}|P_{i}^{l}[j]|=\sum_{j=1}^{m_{i}}|Q_{i}^{l}[j]|=R,$
for $i\in\left\{ 1,\cdots,K_{l}-1\right\} ,$
\item $|P_{K_{l}}^{l}[d_{l}]|=R$ and $|P_{K_{l}}^{l}[k]|=0$ for $k\neq d_{l},$
\item $G_{i}(\bigcup_{k=1}^{m_{i+1}}P_{i+1}^{l}[k],\bigcup_{j=1}^{m_{i}}Q_{i}^{l}[j])$
is a nonsingular matrix for $i\in\left\{ 1,\cdots,K_{l}-1\right\} .$ 
\end{enumerate}
Since $P_{1}^{l}[1],l\in\left\{ 1,\cdots,g\right\} ,$ can be any
subset of $P_{1}[1]$ of size $R,$ we set all subsets $P_{1}^{l}[1],l\in\left\{ 1,\cdots,g\right\} ,$
to be the same subset of $P_{1}[1]$. 

Our design criterion for a multicast coding scheme is that for each
destination $t_{l},l\in\left\{ 1,\cdots,g\right\} ,$ at each layer
$i\in\left\{ 1,\cdots,K_{l}\right\} ,$ the global coding vectors
correponding to the elements of the vectors $\mathbf{y}_{i}(P_{i}^{l}[j])$
for $j\in\left\{ 1,\cdots,m_{i}\right\} $ must be linearly independent
vectors and hence the length--$R$ vector \[
\mathbf{y}_{i}\left(\bigcup_{j=1}^{m_{i}}P_{i}^{l}[j]\right)\]
 can uniquely determine the message vector $\mathbf{w}.$ In other
words we require for each destination $t_{l}$ and each layer $i\in\left\{ 1,\cdots,K_{l}\right\} :$ 
\begin{itemize}
\item Condition ({*}): the matrix  $\boldsymbol{y}_{i}\left(\bigcup_{j=1}^{m_{i}}P_{i}^{l}[j]\right)$
must be nonsingular.
\end{itemize}
The destination node $t_{l}=v_{K_{l}}(d_{l})$ will receive the length--$R$
vector $\mathbf{y}_{K_{l}}(P_{K_{l}}^{l}[d_{l}])=\boldsymbol{y}_{K_{l}}(P_{K_{l}}^{l}[d_{l}])\mathbf{w}.$
Since $\boldsymbol{y}_{K_{l}}(P_{K_{l}}^{l}[d_{l}])$ is a nonsingular
matrix, $t_{l}$ will be able to decode message $\mathbf{w}.$ 

Notice that at each node $v_{i}(j)$ for $i\in\left\{ 2,\cdots,M\right\} $
we only have control over the design of the coding vectors $\boldsymbol{x}_{i}(q)$
for $q\in Q_{i}[j]$ which can be a linear function of the coding
vectors $\left\{ \boldsymbol{y}_{i}(p):p\in P_{i}[j]\right\} .$ The
coding vectors $\boldsymbol{y}_{i}(p)$ for $p\in P_{i}[j]$ are determined
from the coding vectors of the previous layer and matrix $G_{i-1}$.
 In our design we will assign coding vectors layer by layer, starting
from the first layer. At each layer $i$ we fix an arbitrary order
on the elements of the set $Q_{i}$ and assign the coding vectors
$\boldsymbol{x}_{i}(q)$ in this order. 

\textbf{Initialization:} We start from the first layer. Since $P_{1}^{l}[1]$
is the same subset for every $l\in\left\{ 1,\cdots,g\right\} $ we
set $\boldsymbol{y}_{1}(P_{1}^{l}[1])=I_{R\times R}$, i.e., the $R\times R$
identity matrix, and set $\boldsymbol{y}_{1}(P_{1}[1]\setminus P_{1}^{l}[1])=\boldsymbol{0}$
for every $l\in\left\{ 1,\cdots,g\right\} .$ In other words we set
$\mathbf{y}_{1}(P_{1}^{l}[1])=\mathbf{w}$ and $\mathbf{y}_{1}(P_{1}[1]\setminus P_{1}^{l}[1])=\boldsymbol{0}$
for every $l\in\left\{ 1,\cdots,g\right\} $. Therefore condition
({*}) will be satisfied for all destinations in the first layer. 

\textbf{Inductive Step:} We continue our coding construction inductively.
Suppose that the condition ({*}) holds for layer $i$ and for all
destinations $t_{l}=v_{K_{l}}(d_{l})$ with $K_{l}\geq i.$ Next we
will design the coding vectors $\boldsymbol{x}_{i}(q)$ for $q\in Q_{i}$
one by one and in the order of the elements of $Q_{i}$ so that at
the end the condition ({*}) holds for layer $i+1$ and all destinations
$t_{l}=v_{K_{l}}(d_{l})$ with $K_{l}\geq i+1.$

At this step of the algorithm for each destination $t_{l}$ with $K_{l}\geq i+1$
we maintain two matrices. One is the matrix $A_{l}$ which is initially 

\[
A_{l}=\boldsymbol{y}_{i}\left(\bigcup_{j=1}^{m_{i}}P_{i}^{l}[j]\right),\]
and is updated throughout the algorithm. The other matrix is \[
F_{l}=G_{i}(\bigcup_{k=1}^{m_{i+1}}P_{i+1}^{l}[k],Q'_{l}),\]
where initially $Q'_{l}=\bigcup_{j=1}^{m_{i}}Q_{i}^{l}[j]$ and it
is updated throughout the algorithm. Throughout the algorithm we maintain
the invariance that the product $F_{l}A_{l}$ is a nonsingular matrix
for every destination $t_{l}$ with $K_{l}\geq i+1$. We will also
verify that after all of the elements of $Q_{i}$ are processed, for
every destination $t_{l}$ with $K_{l}\geq i+1$ we will have $F_{l}A_{l}=\boldsymbol{y}_{i+1}\left(\bigcup_{j=1}^{m_{i+1}}P_{i+1}^{l}[j]\right)$,
which is sufficient for condition ({*}) to hold at layer $i+1.$ 

$A_{l}$ is initially invertible since condition ({*}) holds for layer
$i.$ Matrix $F_{l}$ is also initially nonsingular by the definition
of a flow to destination $t_{l}$ given in Theorem \ref{thm:unicastflow}.
Therefore the product $F_{l}A_{l}$ is initially nonsingular. Next
we will explain the design of the coding vector $\boldsymbol{x}_{i}(q)$
for $q\in Q_{i}$ and describe the updating process of $F_{l}$ and
$A_{l}$ for every destination $t_{l}$ with $K_{l}\geq i+1$. We
consider two cases:
\begin{enumerate}
\item If $q$ is part of the flow for destination $t_{l},$ i.e., $q\in Q_{i}^{l}[j]$
for some $j\in\left\{ 1,\cdots,m_{i}\right\} ,$ then update matrix
$A_{l}$ by replacing row $\boldsymbol{y}_{i}(p_{l})$ with $\boldsymbol{x}_{i}(q),$
which we will later explain how to design. Here $p_{l}\in P_{i}^{l}[j]$
is the unique element that is matched with $q\in Q_{i}^{l}[j]$ in
the flow for destination $t_{l}$. There is no change needed for matrix
$F_{l}.$
\item If $q$ is not part of the flow for destination $t_{l},$ then update
$A_{l}$ adding a new row $\boldsymbol{x}_{i}(q)$ to it and insert
a column into $F_{l}$ so that the set of column indices grows from
$Q_{l}^{'}$ to $Q'_{l}\cup\left\{ q\right\} $. In this step we place
$\boldsymbol{x}_{i}(q)$ in the row of $A_{l}$ counting from the
top which is the same as the position of the new column $G_{i}(\bigcup_{k=1}^{m_{i+1}}P_{i+1}^{l}[k],\left\{ q\right\} )$
in the updated $F_{l}$ counting from the left.
\end{enumerate}
When we have gone through all of the elements of $Q_{i}$, matrix
$F_{l}$ would be $G_{i}(\bigcup_{k=1}^{m_{i+1}}P_{i+1}^{l}[k],Q_{i})$
and matrix $A_{l}$ would be the matrix $\boldsymbol{x}_{i}(Q_{i}).$
Therefore we have \[
F_{l}A_{l}=G_{i}(\bigcup_{k=1}^{m_{i+1}}P_{i+1}^{l}[k],Q_{i})\boldsymbol{x}_{i}(Q_{i})=\boldsymbol{y}_{i+1}\left(\bigcup_{j=1}^{m_{i+1}}P_{i+1}^{l}[j]\right)\]
where the second equation holds since $G_{i}$ is the transfer matrix
from $\boldsymbol{x}_{i}(Q_{i})=\boldsymbol{x}_{i}$ to $\boldsymbol{y}_{i+1}.$
This equation guarantees that $\boldsymbol{y}_{i+1}\left(\bigcup_{j=1}^{m_{i+1}}P_{i+1}^{l}[j]\right)$
is nonsingular, as desired.

Next we analyze each case and find the condition that $\boldsymbol{x}_{i}(q)$
needs to satisfy in order for $F_{l}A_{l}$ to remain nonsingular:

\subsection*{Analysis of Case 1}

Without loss of generality suppose that $\boldsymbol{x}_{i}(q)$ is
the first row of $A_{l}$ and that matrix $A_{l}$ after the update
is of the form \[
A_{l}=\left[\begin{array}{c}
\boldsymbol{x}_{i}(q)\\
A'_{l}\end{array}\right].\]
Therefore $A_{l}$ before the update is of the form $\left[\begin{array}{c}
\boldsymbol{y}_{i}(p_{l})\\
A'_{l}\end{array}\right].$ We require that the matrix $F_{l}A_{l}$ be nonsingular. We write
\[
F_{l}=\left[\begin{array}{cc}
\boldsymbol{\alpha} & F'_{l}\end{array}\right]\]
 where $\boldsymbol{\alpha}\in\mathbb{F}^{R\times1}$ is the first
column of $F_{l}$. Using standard matrix calculus we can write \begin{align*}
F_{l}A_{l} & =\left[\begin{array}{cc}
\boldsymbol{\alpha} & F'_{l}\end{array}\right]\left[\begin{array}{c}
\boldsymbol{x}_{i}(q)\\
A'_{l}\end{array}\right]\\
 & =\boldsymbol{\alpha}\boldsymbol{x}_{i}(q)+F'_{l}A'_{l}.\end{align*}
 Let us define $H=\left[\begin{array}{cc}
\boldsymbol{\alpha} & F'_{l}\end{array}\right]\left[\begin{array}{c}
\boldsymbol{y}_{i}(p_{l})\\
A'_{l}\end{array}\right],$ which is the matrix $F_{l}A_{l}$ resulting from the previous step
and is nonsingular by the inductive assumption. We can write \[
F'_{l}A'_{l}=H-\boldsymbol{\alpha}\boldsymbol{y}_{i}(p_{l})\]
and therefore \[
F_{l}A_{l}=H+\boldsymbol{\alpha}(\boldsymbol{x}_{i}(q)-\boldsymbol{y}_{i}(p_{l})).\]
For the moment suppose that $F_{l}A_{l}$ is singular. This means
that there exist a non--zero column vector $\boldsymbol{\beta}\in\mathbb{F}^{R\times1}$
with $F_{l}A_{l}\boldsymbol{\beta}=\boldsymbol{0}.$ This implies
that \begin{equation}
H\boldsymbol{\beta}+\boldsymbol{\alpha}(\boldsymbol{x}_{i}(q)-\boldsymbol{y}_{i}(p_{l}))\boldsymbol{\beta}=\boldsymbol{0}.\label{eq:1}\end{equation}
 Since $H$ is a nonsingular matrix, there is a vector $\boldsymbol{\gamma}_{l}$
such that $\boldsymbol{\alpha}=H\boldsymbol{\gamma}_{l}.$ Then (\ref{eq:1})
can be rewritten as \[
H\boldsymbol{\beta}+H\boldsymbol{\gamma}_{l}(\boldsymbol{x}_{i}(q)-\boldsymbol{y}_{i}(p_{l}))\boldsymbol{\beta}=H(\boldsymbol{\beta}+\boldsymbol{\gamma}_{l}(\boldsymbol{x}_{i}(q)-\boldsymbol{y}_{i}(p_{l}))\boldsymbol{\beta})=\boldsymbol{0}.\]
Since $H$ is nonsingular, the identity holds if and only if \[
\boldsymbol{\beta}+\boldsymbol{\gamma}_{l}(\boldsymbol{x}_{i}(q)-\boldsymbol{y}_{i}(p_{l}))\boldsymbol{\beta}=\boldsymbol{0}.\]
If we premultiply the vectors from both sides of the preceding vector
equation by $(\boldsymbol{x}_{i}(q)-\boldsymbol{y}_{i}(p_{l}))$,
we find that\[
(\boldsymbol{x}_{i}(q)-\boldsymbol{y}_{i}(p_{l}))\boldsymbol{\beta}+(\boldsymbol{x}_{i}(q)-\boldsymbol{y}_{i}(p_{l}))\boldsymbol{\gamma}_{l}(\boldsymbol{x}_{i}(q)-\boldsymbol{y}_{i}(p_{l}))\boldsymbol{\beta}=(1+(\boldsymbol{x}_{i}(q)-\boldsymbol{y}_{i}(p_{l}))\boldsymbol{\gamma}_{l})(\boldsymbol{x}_{i}(q)-\boldsymbol{y}_{i}(p_{l}))\boldsymbol{\beta}=0.\]
The expression above is product of two numbers $(1+(\boldsymbol{x}_{i}(q)-\boldsymbol{y}_{i}(p_{l}))\boldsymbol{\gamma}_{l})$
and $(\boldsymbol{x}_{i}(q)-\boldsymbol{y}_{i}(p_{l}))\boldsymbol{\beta}.$
We argue that $(\boldsymbol{x}_{i}(q)-\boldsymbol{y}_{i}(p_{l}))\boldsymbol{\beta}$
is not zero. Observe that if this number was zero, then equation \eqref{eq:1}
and the nonsingularity of $H$ would imply that $H\boldsymbol{\beta}$
and $\boldsymbol{\beta}$ are both zero vectors, contradicting our
assumption that $\boldsymbol{\beta}$ is a non--zero vector. Therefore
\[
1+(\boldsymbol{x}_{i}(q)-\boldsymbol{y}_{i}(p_{l}))\boldsymbol{\gamma}_{l}=0.\]
 This argument implies that for $F_{l}A_{l}$ to be nonsingular it
is sufficient to have the following inequality: \begin{equation}
1+(\boldsymbol{x}_{i}(q)-\boldsymbol{y}_{i}(p_{l}))\boldsymbol{\gamma}_{l}\neq0.\label{eq:inequality1}\end{equation}

\subsection*{Analysis of Case 2}

The analysis is very similar to Case 1. Without loss of generality
assume that the new row is added to the bottom of $A_{l}$ and the
new column is added to the right of $F_{l}$. After the update $A_{l}$
is of the form\[
A_{l}=\left[\begin{array}{c}
A'_{l}\\
\boldsymbol{x}_{i}(q)\end{array}\right].\]
 Here $A'_{l}$ represents matrix $A_{l}$ before the update. Also
matrix $F_{l}$ after the update is of the form \[
F_{l}=\left[\begin{array}{cc}
F'_{l} & \boldsymbol{\alpha}\end{array}\right]\]
 where $\boldsymbol{\alpha}\in\mathbb{F}^{R\times1}$ is the new column
added to $F'_{l}$, which is the matrix $F_{l}$ before the update.
Our inductive assumption implies that $H=F'_{l}A'_{l}$ is nonsingular.
We can write \[
F_{l}A_{l}=H+\boldsymbol{\alpha}\boldsymbol{x}_{i}(q).\]
$F_{l}A_{l}$ is singular if there exists a non--zero vector $\boldsymbol{\beta}$
such that \[
F_{l}A_{l}\boldsymbol{\beta}=H\boldsymbol{\beta}+\boldsymbol{\alpha}\boldsymbol{x}_{i}(q)\boldsymbol{\beta}=\boldsymbol{0}.\]
 Since $H$ is nonsingular, there exists a vector $\boldsymbol{\gamma}_{l}$
such that $\boldsymbol{\alpha}=H\boldsymbol{\gamma}_{l}.$ Therefore
$FA_{l}$ is singular if there exists a non--zero vector $\boldsymbol{\beta}$
for which\begin{equation}
H\boldsymbol{\beta}+H\boldsymbol{\gamma}_{l}\boldsymbol{x}_{i}(q)\boldsymbol{\beta}=H(\boldsymbol{\beta}+\boldsymbol{\gamma}_{l}\boldsymbol{x}_{i}(q)\boldsymbol{\beta})=\boldsymbol{0}.\label{eq:equation3}\end{equation}
 Since $H$ is nonsingular, (\ref{eq:equation3}) implies \[
\boldsymbol{\beta}+\boldsymbol{\gamma}_{l}\boldsymbol{x}_{i}(q)\boldsymbol{\beta}=\boldsymbol{0}.\]
 If we premultiply both sides of the preceding equation by $\boldsymbol{x}_{i}(q)$
we obtain \[
\boldsymbol{x}_{i}(q)\boldsymbol{\beta}+\boldsymbol{x}_{i}(q)\boldsymbol{\gamma}_{l}\boldsymbol{x}_{i}(q)\boldsymbol{\beta}=\boldsymbol{x}_{i}(q)\boldsymbol{\beta}(1+\boldsymbol{x}_{i}(q)\boldsymbol{\gamma}_{l})=0.\]
 The previous equality holds if either $\boldsymbol{x}_{i}(q)\boldsymbol{\beta}=0$
or if $(1+\boldsymbol{x}_{i}(q)\boldsymbol{\gamma}_{l})=0.$ If $\boldsymbol{x}_{i}(q)\boldsymbol{\beta}=0$
then by \eqref{eq:equation3} $H\boldsymbol{\beta}=\boldsymbol{0}$,
which, together with the invertibility of $H$, implies that $\boldsymbol{\beta}=\boldsymbol{0}$.
But $\boldsymbol{\beta}\neq\boldsymbol{0}$ by assumption. Therefore
if $F_{l}A_{l}$ is a singular matrix, we have \[
1+\boldsymbol{x}_{i}(q)\boldsymbol{\gamma}_{l}=0.\]
 The preceding argument implies that $F_{l}A_{l}$ is nonsingular
if \begin{equation}
1+\boldsymbol{x}_{i}(q)\boldsymbol{\gamma}_{l}\neq0.\label{eq:inequality2}\end{equation}

\subsection*{A randomized algorithm and the existence of a solution}

Let us summarize the analysis up to this point. The coding vectors
$\boldsymbol{x}_{i}(q),q\in Q_{i}$, can be assigned in a way that
meet our requirements if\[
\tau\triangleq\prod_{t_{l}:q\in Q_{i}^{l}[j],j\in\left\{ 1,\cdots,m_{i}\right\} }\left(1+(\boldsymbol{x}_{i}(q)-\boldsymbol{y}_{i}(p_{l}))\boldsymbol{\gamma}_{l}\right)\prod_{t_{l}:q\notin Q_{i}^{l}[j],j\in\left\{ 1,\cdots,m_{i}\right\} }\left(1+\boldsymbol{x}_{i}(q)\boldsymbol{\gamma}_{l}\right)\neq0.\]
In the preceding equation $t_{l}$ is restricted to the destinations
for which $K_{l}\geq i+1,$ and the vectors $\boldsymbol{\gamma}_{l}$
and $\boldsymbol{y}_{i}(p_{l})$ are specified in the analyses of
Cases 1 and 2.

One other constraint is that $\boldsymbol{x}_{i}(q)$ for $q\in Q_{i}[j]$
can only be a linear combination of the vectors $\left\{ \boldsymbol{y}_{i}(p):p\in P_{i}[j]\right\} .$
Let us choose each $\boldsymbol{x}_{i}(q)$ to be a random linear
combination of the elements of $\left\{ \boldsymbol{y}_{i}(p):p\in P_{i}[j]\right\} $
where the coefficient of each $\boldsymbol{y}_{i}(p)$ is randomly
and independently chosen from the uniform distribution over the field
$\mathbb{F}.$ For each destination $t_{l}$ with $K_{l}\geq i+1$
define $\phi_{l}$ as the event that the corresponding term in the
product above is zero. Then we have \[
\Pr(\tau=0)=\Pr(\bigvee_{t_{l}:K_{l}\geq i+1}\phi_{l})\leq\sum_{t_{l}:K_{l}\ge i+1}\Pr(\phi_{l}).\]
 Suppose that $q\in Q_{i}[j]$ and $\boldsymbol{x}_{i}(q)=\sum_{p\in P_{i}[j]}\theta_{p}\boldsymbol{y}_{i}(p).$
Now consider a destination $t_{l}$ with $K_{l}\geq i+1$. If $q\in Q_{i}^{l}[j]$
and $p_{l}\in P_{i}^{l}[j]$ is matched with $q$, we need to have
$1+(\boldsymbol{x}_{i}(q)-\boldsymbol{y}_{i}(p_{l}))\boldsymbol{\gamma}_{l}\neq0$.
There exist $\omega_{0}\in\mathbb{F},\omega_{p}\in\mathbb{F},p\in P_{i}[j],$
which are determined by $\boldsymbol{y}_{i}(p)$ and $\boldsymbol{\gamma}_{l}$
and satisfy \[
1+(\boldsymbol{x}_{i}(q)-\boldsymbol{y}_{i}(p_{l}))\boldsymbol{\gamma}_{l}=\omega_{0}+\sum_{p\in P_{i}[j]}\omega_{p}\theta_{p}\]
 There are two cases to consider. First, if $\omega_{p}=0$ for all
$p\in P_{i}[j],$ then $\omega_{0}+\sum_{p\in P_{i}[j]}\omega_{p}\theta_{p}=\omega_{0}$
is a constant independent of $\theta_{p},p\in P_{i}[j]$. Furthermore
by setting $\theta_{p_{l}}=1$ and $\theta_{p}=0$ for $p\in P_{i}[j]$
and $p\neq p_{l}$ so that $\boldsymbol{x}_{i}(q)=\boldsymbol{y}_{i}(p_{l}),$
we find that  \[
\omega_{0}=1+(\boldsymbol{x}_{i}(q)-\boldsymbol{y}_{i}(p_{l}))\boldsymbol{\gamma}_{l}=1.\]
Therefore in this case $\Pr(\phi_{l})=0.$ Next if there exists some
$p\in P_{i}[j]$ for which $\omega_{p}\neq0$ then $\omega_{0}+\sum_{p\in P_{i}[j]}\omega_{p}\theta_{p}$
depends on $\theta_{p},p\in P_{i}[j].$ Since $\theta_{p},p\in P_{i}[j]$,
are uniformly distributed random variables over $\mathbb{F},$ $\omega_{0}+\sum_{p\in P_{i}[j]}\omega_{p}\theta_{p}$
is likewise uniformly distributed over $\mathbb{F}$. In this case
$\Pr(\phi_{l})=\frac{1}{|\mathbb{F}|}.$ 

Next suppose that $q\notin Q_{i}^{l}[j]$. Here we need to have $1+\boldsymbol{x}_{i}(q)\boldsymbol{\gamma}_{l}\neq0$.
Following the preceding argument, there exist $\omega_{0}\in\mathbb{F},\omega_{p}\in\mathbb{F},p\in P_{i}[j],$
which are determined by $\boldsymbol{y}_{i}(p)$ and $\boldsymbol{\gamma}_{l}$
and satisfy \[
1+\boldsymbol{x}_{i}(q)\boldsymbol{\gamma}_{l}=\omega_{0}+\sum_{p\in P_{i}[j]}\omega_{p}\theta_{p}.\]
If $\omega_{p}=0$ for all $p\in P_{i}[j],$ then $\omega_{0}+\sum_{p\in P_{i}[j]}\omega_{p}\theta_{p}=\omega_{0}$
is a constant independent of $\theta_{p},p\in P_{i}[j]$. By setting
$\theta_{p}=0$ for all $p\in P_{i}[j]$ so that $\boldsymbol{x}_{i}(q)=\boldsymbol{0},$
we obtain $\omega_{0}=1+\boldsymbol{x}_{i}(q)\boldsymbol{\gamma}_{l}=1$,
and $\Pr(\phi_{l})=0.$ If there is some $p\in P_{i}[j]$ for which
$\omega_{p}\neq0,$ then an analogous argument to our earlier one
implies that $\Pr(\phi_{l})=\frac{1}{|\mathbb{F}|}.$ 

As a result, for any destination $t_{l}$ with $K_{l}\ge i+1,$ we
have $\Pr(\phi_{l})\leq\frac{1}{|\mathbb{F}|}.$ Therefore \[
\Pr(\tau=0)\leq\sum_{t_{l}:K_{l}\ge i+1}\Pr(\phi_{l})\leq\frac{g}{|\mathbb{F}|}.\]
 Since we are interested in the event that $\tau\neq0$, we have \[
\Pr(\tau\neq0)\geq1-\frac{g}{|\mathbb{F}|}.\]
 Therefore if $|\mathbb{F}|>g,$ then $\Pr(\tau\neq0)>0$ and there
is at least one valid solution for $\boldsymbol{x}_{i}(q).$ This
also yields a randomized algorithm with probability of success of
at least $1-\frac{g}{|\mathbb{F}|}.$ If we take the size of the field
to be $|\mathbb{F}|\geq2g$ then the probability of success will be
at least $1-\frac{g}{2g}=\frac{1}{2}$ for each $q\in Q_{i}.$

\subsection*{A deterministic polynomial time algorithm}

We next explain a deterministic algorithm with polynomial running
time for the finding vectors $\boldsymbol{x}_{i}(q),q\in Q_{i}[j].$
For each $q\in Q_{i}$ we seek a vector $\boldsymbol{u}=\boldsymbol{x}_{i}(q)$
which is a linear combination of the vectors in $\left\{ \boldsymbol{y}_{i}(p):p\in P_{i}[j]\right\} $
such that for any destination $t_{l}$ with $K_{l}\geq i+1$, if $q\in Q_{i}^{l}[j]$
and $p_{l}\in P_{i}^{l}[j]$ is matched with $q$, then $1+(\boldsymbol{u}-\boldsymbol{y}_{i}(p_{l}))\boldsymbol{\gamma}_{l}\neq0$,
and if $q\notin Q_{i}^{l}[j]$ then $1+\boldsymbol{u}\boldsymbol{\gamma}_{l}\neq0$. 

Define the subset of indices of destinations $W$ as\[
W=\left\{ l\in\left\{ 1,\cdots,g\right\} :K_{l}\geq i+1,q\in Q_{i}^{l}[j]\mbox{ for some }j\in\left\{ 1,\cdots,m_{i}\right\} ,\boldsymbol{y}_{i}(p_{l})\boldsymbol{\gamma}_{l}\neq0\right\} .\]
 We can write the conditions that $\boldsymbol{u}$ needs to satisfy
as $1+(\boldsymbol{u}-\boldsymbol{y}_{i}(p_{l}))\boldsymbol{\gamma}_{l}\neq0$
for $l\in W$ and $1+\boldsymbol{u}\boldsymbol{\gamma}_{l}\neq0$
for $l\notin W$ and $K_{l}\geq i+1.$ Next we use \cite[Lemma 8]{jaggi}:
\begin{lem}
Let $n\leq|\mathbb{F}|.$ Let $\boldsymbol{a}_{1},\cdots,\boldsymbol{a}_{n}\in\mathbb{F}^{1\times R}$
and $\boldsymbol{b}_{1},\cdots,\boldsymbol{b}_{n}\in\mathbb{F}^{R\times1}$
with $\boldsymbol{a}_{i}\boldsymbol{b}_{i}\neq0,i\in\left\{ 1,\cdots,n\right\} .$
There exists a linear combination $\boldsymbol{c}$ of $\boldsymbol{a}_{1},\cdots,\boldsymbol{a}_{n}$
such that $\boldsymbol{c}\boldsymbol{b}_{i}\neq0,i\in\left\{ 1,\cdots,n\right\} .$
Such a vector $\boldsymbol{c}$ can be found in time $O(n^{2}R).$ 
\end{lem}
If we are given the set of vectors $\boldsymbol{\gamma}_{l}$, then
it takes $O(gR)$ steps to form the set $W.$ Then by applying the
preceding lemma, if $g\leq|\mathbb{F}|,$ we can find a vector $\boldsymbol{w}\in\mathbb{F}^{1\times R}$
such that $\boldsymbol{w}$ is a linear combination of the vectors
in $\left\{ \boldsymbol{y}_{i}(p_{l}):l\in W\right\} $ and for every
$l\in W,$ we have that $\boldsymbol{w}\boldsymbol{\gamma}_{l}\neq0.$
Furthermore vector $\boldsymbol{w}$ can be found in time $O(g^{2}R).$
By adding the time $O(gR)$ needed to produce set $W,$ we need a
total time of $O(g^{2}R+gR)=O(g^{2}R)$ to find vector $\boldsymbol{w}$.
Next we let $\boldsymbol{u}=\sigma\boldsymbol{w}$ for some $\sigma\in\mathbb{F}.$
We show that an appropriate value of $\sigma$ exists such that $\boldsymbol{u}$
satisfies all of the constraints. 

For $l\in W,$ we need to have $1+(\sigma\boldsymbol{w}-\boldsymbol{y}_{i}(p_{l}))\boldsymbol{\gamma}_{l}\neq0.$
Therefore \begin{equation}
\sigma\neq\frac{\boldsymbol{y}_{i}(p_{l})\boldsymbol{\gamma}_{l}-1}{\boldsymbol{w}\boldsymbol{\gamma}_{l}}.\label{eq:constraintW}\end{equation}
 For $l\notin W$ and $K_{l}\geq i+1$ we need to have $1+\sigma\boldsymbol{w}\boldsymbol{\gamma}_{l}\neq0.$
If $\boldsymbol{w}\boldsymbol{\gamma}_{l}=0$ then this condition
is fulfiled for all values of $\sigma.$ Otherwise we need to have
\begin{equation}
\sigma\neq\frac{-1}{\boldsymbol{w}\boldsymbol{\gamma}_{l}}.\label{eq:constraintnotW}\end{equation}
 There are at most $g$ constraints of the form \eqref{eq:constraintW}
and \eqref{eq:constraintnotW} on $\sigma.$ Therefore if the size
of field $\mathbb{F}$ is greater than the number of destinations
$g$, this deterministic approach will find at least one $\sigma$
that is not in the discriminating set by at most considering $g$
elements of $\mathbb{F}$. Therefore the total complexity of finding
an appropriate value of $\sigma$ is $O(g)$ and the total complexity
of finding vector $\boldsymbol{u}$ is $O(g+g^{2}R)=O(g^{2}R).$ 

To find the overall complexity of finding the vector $\boldsymbol{x}_{i}(q),$
we need to evaluate the complexity of finding vector $\boldsymbol{\gamma}_{l}$
for every $l\in\left\{ 1,\cdots,g\right\} $ with $K_{l}\geq i+1$.
From the analysis of Cases 1 and 2, $\boldsymbol{\gamma}_{l}=H^{-1}\boldsymbol{\alpha}$,
where matrix $H$ is $F_{l}A_{l}$ from the previous step of the algorithm.
Since matrix $F_{l}$ has size $R\times L$ and matrix $A_{l}$ has
size $L\times R$ for some $R\leq L\leq|Q_{i}|$, computing $H$ needs
$O(R|Q_{i}|)$ operations. Evaluating $H^{-1}$ also needs $O(R^{3})$
steps and so there are a total of $O(R|Q_{i}|+R^{3})$ operations
for evaluating $\boldsymbol{\gamma}_{l}.$ Since there are at most
$g$ different $l\in\left\{ 1,\cdots,g\right\} $ with $K_{l}\geq i+1$,
we will have $O(gR|Q_{i}|+gR^{3})$ as the total complexity of evaluating
different values of $\boldsymbol{\gamma}_{l}$ for any specific $q\in Q_{i}.$
Therefore the total complexity of evaluating $\boldsymbol{x}_{i}(q)$
will be $O(gR|Q_{i}|+gR^{3}+g^{2}R).$ 

Let us assume that the number of nodes $m_{i}$ at each layer $i\in\left\{ 1,\cdots,M\right\} $
is at most $m.$ Furthermore assume that the size of transmitted and
received signals at each node is at most $r.$ Therefore the total
complexity of evaluating each $\boldsymbol{x}_{i}(q)$ will be $O(gRmr+gR^{3}+g^{2}R).$
Since there are at most $mMr$ different $\boldsymbol{x}_{i}(q)$
to be evaluated, if we assume that the unicast flows from source to
each destination is provided, the total complexity of our algorithm
is $O(gRm^{2}Mr^{2}+gR^{3}mMr+g^{2}RmMr)=O(gRmMr(mr+R^{2}+g)).$

The complexity of computing a unicast flow to a destination by the
algorithm given in \cite{goemans} is $O(M(mr)^{3}\log mr).$ Since
we have $g$ destinations, the total complexity of computing the unicast
flows will be $O(gM(mr)^{3}\log mr).$ If we add this running time
to the running time of our algorithm, the total running time will
be $O(gmMr(mrR+R^{3}+gR+(mr)^{2}\log mr)).$ We can compare it to
the running time of the algoithm given in \cite{ebrahimi} which is
$O(g(r^{2}mM+R)^{3}\log(r^{2}mM+R)+r^{2}mM(r^{2}mM+R)^{2}+(g\log gRM)^{3})$
and see that our algorithm is considerably faster.

\subsection*{Number of network uses to achieve capacity}

We have shown that it is sufficient for the size of the field of operation
$\mathbb{F}$ of the LDRN to be greater than $g$ to guarantee the
existence of a multicast coding solution. In general however, the
network operates over some fixed field which is usually $\mathbb{F}_{p}$
for some prime number $p.$ In order to achieve a greater field size,
we will use multiple rounds of the network. Here we will argue that
if we use the network for $k$ rounds, it is equivalent to an LDRN
with field of operation $\mathbb{F}=\mathbb{F}_{p}^{k}.$ This implies
that in order to have a field size at least $g+1,$ it is sufficient
to use the network for $k=\left\lceil \log_{p}(g+1)\right\rceil $
rounds. This is an improvement over the number of rounds that is needed
for the algorithm of \cite{ebrahimi} which is $k\approxeq\log_{p}(g(\log_{p}g-1)RM).$

Suppose that the network is used for $k$ rounds and we use the superscript
$t\in\left\{ 0,\cdots,k-1\right\} $ to denote the time index that
a vector is received or sent. For each $i\in\left\{ 1,\cdots,M-1\right\} $
we have\[
\mathbf{y}_{i+1}^{t}=G_{i}\mathbf{x}_{i}^{t},\qquad t\in\left\{ 0,\cdots,k-1\right\} \]
Observe we can use a dummy variable $D$ as the unit delay operator
and represent the preceding $k$ equations as a single equation\[
\sum_{t=0}^{k-1}\mathbf{y}_{i+1}^{t}D^{t}=G_{i}\sum_{t=0}^{k-1}\mathbf{x}_{i}^{t}D^{t}.\]
Next, notice that $\sum_{t=0}^{k-1}\mathbf{y}_{i+1}^{t}D^{t}$ and
$\sum_{t=0}^{k-1}\mathbf{x}_{i}^{t}D^{t}$ can be regarded as new
vectors in the extension field $\mathbb{F}_{p}^{k}$ and we can assume
that the network is operating in the extension field $\mathbb{F}_{p}^{k}$.
Since the transfer matrix between the layers $i$ and $i+1$ is still
$G_{i}$ and has not changed in the new field, the existence of the
unicast flow over the original field implies the existence of flow
over the extended field. Therefore our analysis is valid over any
field $\mathbb{F}_{p}^{k}.$

\end{document}